\documentclass[aps,pra,twocolumn,floats,showpacs,floatfix]{revtex4-1}
\usepackage{color}
\usepackage{bm}
\usepackage{latexsym}
\usepackage{amsmath}
\usepackage{amssymb}
\usepackage[pdftex]{graphicx}

\def\ln{\mbox{ln}}

\begin{document}

\newcommand{\cc}{{\bf\Large C }}
\newcommand{\half}{\mbox{\small $\frac{1}{2}$}}
\newcommand{\sinc}{\mbox{sinc}}
\newcommand{\trc}{\mbox{trace}}
\newcommand{\intt}{\int\!\!\!\!\int }
\newcommand{\ointt}{\int\!\!\!\!\int\!\!\!\!\!\circ\ }
\newcommand{\ar}{\mathsf r}
\newcommand{\bmsf}[1]{\bm{\mathsf{#1}}}
\newcommand{\dd}[1]{\:\mbox{d}#1}

\newcommand{\ham}{\mathcal{H}}
\newcommand{\PT}{\mathcal{PT}}
\newcommand{\const}{\mbox{const.}}
\newcommand{\EPS} {\mbox{\LARGE $\epsilon$}}

\newcommand{\hide}[1]{}
\newcommand{\tbox}[1]{\mbox{\tiny #1}}
\newcommand{\mbf}[1]{{\mathbf #1}}

\newcommand{\im}[1]{\mbox{Im}\left\{#1\right\}}
\newcommand{\re}[1]{\mbox{Re}\left\{#1\right\}}
\newcommand{\abs}[1]{\left|#1\right|}
\newcommand{\bra}[1]{\left\langle #1\right|}
\newcommand{\ket}[1]{\left|#1\right\rangle }
\newcommand{\eexp}{\mbox{e}^}

\newcommand{\eq}[1]{Eq.~(\ref{eq:#1})}
\newcommand{\fig}[1]{Fig.~\ref{fig:#1}}
\newcommand{\eos}{\,.}
\definecolor{red}{rgb}{1,0.0,0.0}


\title{Unidirectional Nonlinear $\PT$-symmetric Optical Structures}
\author{Hamidreza Ramezani$^{1}$ and Tsampikos Kottos$^{1,2}$}
\affiliation{
$^1$Department of Physics, Wesleyan University, Middletown, Connecticut 06459, USA\\
$^2$Max-Planck-Institute for Dynamics and Self-Organization, 37073G\"ottingen, Germany
}
\author{Ramy El-Ganainy and Demetrios N. Christodoulides}
\affiliation{
College of Optics \& Photonics-CREOL, University of Central Florida, Orlando, Florida 32816, USA
}

\date{\today}

\begin{abstract}
We show that non-linear optical structures involving a balanced gain-loss profile, can act as
unidirectional optical valves. This is made possible by exploiting the interplay between the 
fundamental symmetries of parity (${\cal P}$) and time (${\cal T}$), with optical nonlinear
effects. This novel unidirectional dynamics is specifically demonstrated for the case of an 
integrable $\PT$-symmetric nonlinear system.
\end{abstract}

\pacs{42.25.Bs, 11.30.Er, 42.82.Et}
\maketitle

\section{Introduction}

Transport phenomena and in particular directed transport are at the heart of many fundamental problems 
in physics, chemistry and biology \cite{K05}. At the same time they are also of great relevance to 
technological applications based on a variety of transport-based devices such as rectifiers, pumps, 
particle separators, molecular switches, and electronic diodes and transistors. Of special interest is 
the realization of novel classes of integrated photonic devices that allow one-directional flow of 
information-e.g. optical isolators \cite{ST91}. Currently, such unidirectional elements rely mainly
on the Faraday effect, where external magnetic fields are used to break the space-time symmetry. 
This in general requires materials with appreciable Verdet constants-typically non compatible with 
light-emitting wafers \cite{ST91}. To anticipate these problems, alternative proposals for the creation 
of optical diodes and isolators have been suggested in recent years. Some representative examples 
include the creation of optical diodes based on asymmetric nonlinear absorption \cite{PAYR07}, second 
harmonic generation in asymmetric waveguides \cite{GAPF01}, nonlinear photonic crystals \cite{SDBB94}, 
and photonic quasi-crystals and molecules \cite{B08}.

In this paper, we propose a new mechanism for unidirectional optical transport based on configurations
involving nonlinear optical materials with Parity (${\cal P}$) and Time (${\cal T}$)-reflection. This is 
possible by judiciously interleaving gain and loss regions, in such a way that the (complex) refractive 
index $n(x)=n_R(x) + i \gamma n_I(x)$ profile satisfies the condition $n^*(-x)=n(x)$. A first experimental 
realization of such (linear) arrangements has been recently reported in Refs. \cite{GSDMVASC09,Secret} 
where a ${\cal PT}$ dual coupled structure was fabricated and the beam dynamics was investigated. Here 
we show that the interplay of non-reciprocal dynamics arising from ${\cal PT}$-symmetry \cite{Secret}, 
and self-trapping phenomena associated with Kerr nonlinearities \cite{J82}, can mold the flow of light 
in a surprising way. Such novel directed dynamics can be exploited in the realization of a new generation 
of optical isolators or diodes. 

Even though the validity of our arguments can be demonstrated for a 
variety of non-linear ${\cal PT}$-configurations, below, we will highlight its basic principles, using the 
simplest possible arrangement, consisting of two $\PT$-coupled waveguide elements with Kerr nonlinearity of 
strength $\chi$. Each of the waveguides is single-moded-- one providing gain and the other an equal 
amount of loss (see Fig.~\ref{fig1}). We have obtained the phase diagram in the $\chi-\gamma$ plane for 
which our system acts as an optical diode, and we have identified the minimum propagation length 
needed, in order to achieve this unidirectional functionality. Detail numerical simulations support our 
theoretical predictions. 

This paper is structured as follows. In Sec. II an overview of the linear ${\cal PT}$-symmetric 
dimer is presented. The nonlinear ${\cal PT}$-symmetric dimer will be introduced in Sec. III, where the 
equations of motion are given in terms of Stokes parameters. In subsection III.A, we present both our
theoretical and numerical results on the dynamics of the non-linear ${\cal PT}$-symmetric dimer. In subsection 
III.B we calculate the critical value of the non-linearity for which diode action is possible. Finally we 
will draw our conclusions in Sec. IV.

\section{Linear ${\cal PT}$-symmetric dimer: An overview}

In this section we will briefly review the basic properties of the linear ${\cal PT}$-symmetric dimer 
\cite{GSDMVASC09,Secret,Kottos}. In integrated optics this simple ${\cal PT}$ element can be realized
in the form of a coupled system, with only one of the two parallel channels being optically pumped to 
provide gain $\gamma$ for the guided light, whereas the neighbour arm experiences equal amount of loss
(see Fig. 1).
Under these conditions, and by using the coupled-mode approach, the optical-field dynamics in the two
coupled waveguides are described by the following set of equations:
\begin{equation}
\label{eq1a}
\begin{array}{cccc}
i{d\psi_1 \over dz} +\psi_{2} - i \gamma \psi_1 &=&0;&\quad (a) \\
i{d\psi_2 \over dz} +\psi_{1} + i \gamma \psi_2 &=&0;&\quad (b) \\
\end{array}
\end{equation}
where $\psi_{1,2}$ are modal electric field amplitudes in the amplifying (Eq.~\ref{eq1a}a) and lossy (Eq.~
\ref{eq1a}b) waveguide channels, $z$ represents a dimensionless propagation distance- normalized in units 
of coupling lengths, and $\gamma$ is a scaled gain/loss coefficient, also normalized to the coupling strength. 

The Hamiltonian corresponding to the linear problem of Eq. (\ref{eq1a}), is written as:
\begin{equation}
\label{PTH}
{\cal H}=\left(
\begin{array}{cc}
i\gamma & -1\\
-1&-i\gamma 
\end{array}
\right)
\end{equation}
and commutes with the combined ${\cal PT}$ operator. A surprising result associated with this class of problems 
is the possibility that such a ${\cal PT}$ symmetric Hamiltonian ${\cal H}$ can have an entirely real energy 
spectrum, despite the fact that, in general, they are non-Hermitian 
\cite{Secret,Kottos,BB98,BBM98,BFKS09,BM08,GMCM07,MGCM08,MMGC08,KGM08,L09,GSDMVASC09}. For the specific example
of the non-hermitian Hamiltonian of Eq. (\ref{PTH}), a direct diagonalization gives the following 
set of eigenvalues:
\begin{equation}
\label{eival}
\lambda_{\pm}=\pm \sqrt{1-\gamma^2}
\end{equation}
which are real as long as the gain/loss parameter $\gamma$ is smaller than some critical value, $\gamma_{\cal 
PT}=1$ ({\it exact} ${\cal PT}$-symmetric phase). As the gain/loss parameter $\gamma$ increases above $\gamma_{
\cal PT}$, the eigenvalues becomes complex ({\it broken} ${\cal PT}$-symmetric phase). The corresponding eigenvectors 
of the Hamiltonian Eq. (\ref{PTH}) are
\begin{eqnarray}
\label{evec}
|+\rangle = 
\left(
\begin{array}{c}
e^{i{\alpha\over 2}}\\
e^{-i{\alpha\over 2}}
\end{array} 
\right)
,& 
|-\rangle = 
\left(
\begin{array}{c}
ie^{-i{\alpha\over 2}}\\
-ie^{i{\alpha\over 2}}
\end{array}  
\right);\,\, \sin\alpha=\gamma
\end{eqnarray}

In the exact ${\cal PT}$-symmetric phase, both the ${\cal H}$ and ${\cal PT}$ operators share the same set of eigenvectors. 
In this regime, the mode intensity is symmetric with respect to the mirror axis of the two waveguides. As $\gamma$ 
increases above $\gamma_{\cal PT}$ the eigenfunctions of ${\cal H}$ cease to be 
eigenfunctions of the ${\cal PT}$-operator, despite the fact that ${\cal H}$ and the ${\cal PT}$-operator still commute. This 
happens because the ${\cal PT}$-operator is anti-linear, and thus the eigenstates of ${\cal H}$ may or may not be eigenstates 
of ${\cal PT}$. In the broken ${\cal PT}$-symmetric phase, the spacial distribution of 
the modes is asymmetric, one of them living predominantly in the amplifying waveguide and the other in the lossy one.  At the 
phase-transition point $\gamma = \gamma_{\cal PT}$ the two eigenfunctions and their corresponding eigenvalues coalesce 
leading to an ``exceptional" point singularity \cite{B03}.  

The beam dynamics associated with Eq. (\ref{eq1a}) were investigated theoretically in Refs. \cite{GSDMVASC09,Kottos,BB98} 
while direct measurements were performed in \cite{Secret,GSDMVASC09}. These authors recognized that as the gain/loss 
parameter $\gamma$ reaches $\gamma_{\cal PT}$, the total beam power starts growing exponentially, while for $\gamma < 
\gamma_{\cal PT}$ power oscillations are observed (see Figs. 1b,c). The most dramatic effect in the beam evolution is the 
appearance of non-reciprocal wave propagation (see Figs. 1c-f). Specifically, the beam propagation pattern differs depending on whether 
the initial excitation is on the left or right waveguide. This is contrasted with the $\gamma=0$ case (Fig. 1a,b), where 
the beam propagation is insensitive to the initial condition.

\begin{figure}[h]
\includegraphics[width=1.0\columnwidth,keepaspectratio,clip]{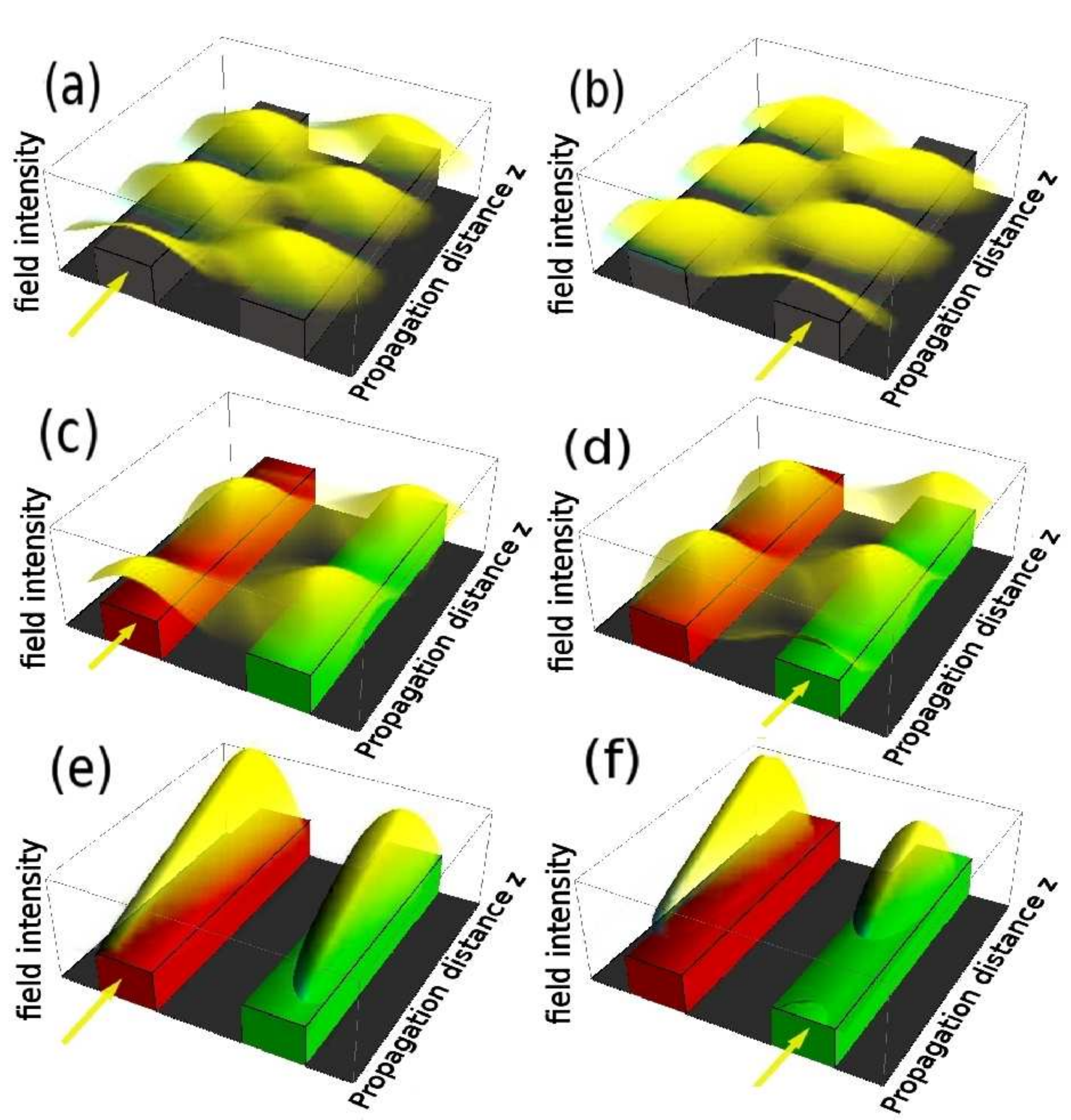}
 \caption{
   \label{fig1_l} 
   (color online) Beam propagation in two coupled linear waveguides. For the parameters of the simulation 
(we use normalized coupling units), the spontaneous ${\cal PT}$-breaking take place at $\gamma_{\cal PT}=1$. 
In all cases, left (right) panels correspond to an initial excitation at the left (right) channel. (a,b) A 
passive system corresponding to $\gamma=0$. The propagation is reciprocal; 
(c,d) $\gamma=0.4\gamma_{\cal PT}$ corresponding to the exact ${\cal PT}$-phase. A non-reciprocal beam propagation
is evident. Although the dynamics is non-Hermitian, the evolution is ``pseudo-unitary" and the total beam power 
remains bounded; (e,f) For $\gamma=1.5\gamma_{\cal PT}$, beam power grows exponentially (vertical scale is logarithmic)
in both waveguides, while the beam propagation is again non-reciprocal with respect to the mirror axis of the 
two waveguides. Waveguides are color-coded, indicating balanced gain (red) and loss (green) regions. Grey-color
waveguides indicate a passive ($\gamma=0$) system. 
   }
\end{figure}

\section{Non-linear ${\cal PT}$-symmetric dimer}

We begin our analysis by providing the mathematical model that describes optical wave propagation in a Kerr 
nonlinear ${\cal PT}$ symmetric coupled dual waveguide arrangement (see Fig.~\ref{fig1}). The two modal 
field amplitudes are governed by the evolution equations:
\begin{equation}
\label{eq1}
\begin{array}{cccc}
i{d\psi_1 \over dz} +\psi_{2} - i \gamma \psi_1 + \chi |\psi_1|^2\psi_1 &=&0;&\quad (a) \\
i{d\psi_2 \over dz} +\psi_{1} + i \gamma \psi_2 + \chi |\psi_2|^2\psi_2 &=&0;&\quad (b) \\
\end{array}
\end{equation}
where $\chi$ is the strength of the Kerr-nonlinearity. 

Equations (\ref{eq1}) can be rewritten in terms of the (real) Stokes parameters  $S_i= \psi^{\dagger} {\hat 
\sigma}_i \psi$, where ${\hat \sigma}_i (i=0,1,2,3)$ denote the Pauli spin matrices \cite{H87}. In this 
representation, the total field intensity is given by $S_0=|\psi_1|^2+|\psi_2|^2$, $S_3= |\psi_1|^2- |\psi_2|^2$ 
is the intensity imbalance between the two waveguides, while $S_1=\psi_1^*\psi_2+\psi_1\psi_2^*$ and 
$S_2=i (\psi_1\psi_2^*-\psi_1^*\psi_2)$. In this representation Eqs. (\ref{eq1}) take the form:
\begin{equation}
\label{eq5}
{dS_0\over dz} = {\overrightarrow E} \cdot {\overrightarrow S};\quad 
{d{\overrightarrow S}\over dz} = S_0{\overrightarrow E} + {\overrightarrow S} \times {\overrightarrow B}
\end{equation}
where we have introduced the two real vectors ${\overrightarrow E} =(0,0,2\gamma)$ and ${\overrightarrow B} 
=(2,0,\chi S_3)$, and the $3$-dimensional Stokes vector ${\overrightarrow S}\equiv (S_1,S_2,S_3)$. We note
that the condition $S_0^2-{\overrightarrow S}\cdot {\overrightarrow S}=0$ is always satisfied. It is worth 
mentioning that Eqs. (\ref{eq5}), are identical to the equation of motion of a relativistic negatively
-charged particle with zero mass, in a pseudo-electromagnetic field $({\overrightarrow E}, {\overrightarrow B})$, 
where $(S_0, {\overrightarrow S})$ represents the energy and $3-$dimensional momentum of the particle, 
while the propagation distance $z$ has the role of the time. 

Nonlinear $\PT$-symmetric optical coupled systems can be realistically synthesized on semiconductor wafers-
known for their high Kerr-like nonlinearities \cite{DCB87}. As in Ref. \cite{Secret}, coupling lengths 
as low as $L_c=1mm$ can be obtained, in which case a gain/loss level below $\pm 30cm^{-1}$ (readily 
available in such materials) will suffice to keep the arrangement in the $\PT$ phase. In addition, 
critical switching ($\chi \sim 1$) can also occur at milliwatt power levels in multi-quantum well 
configurations.   

\subsection{Dynamics}

\begin{figure}[h]
\includegraphics[width=1.0\columnwidth,keepaspectratio,clip]{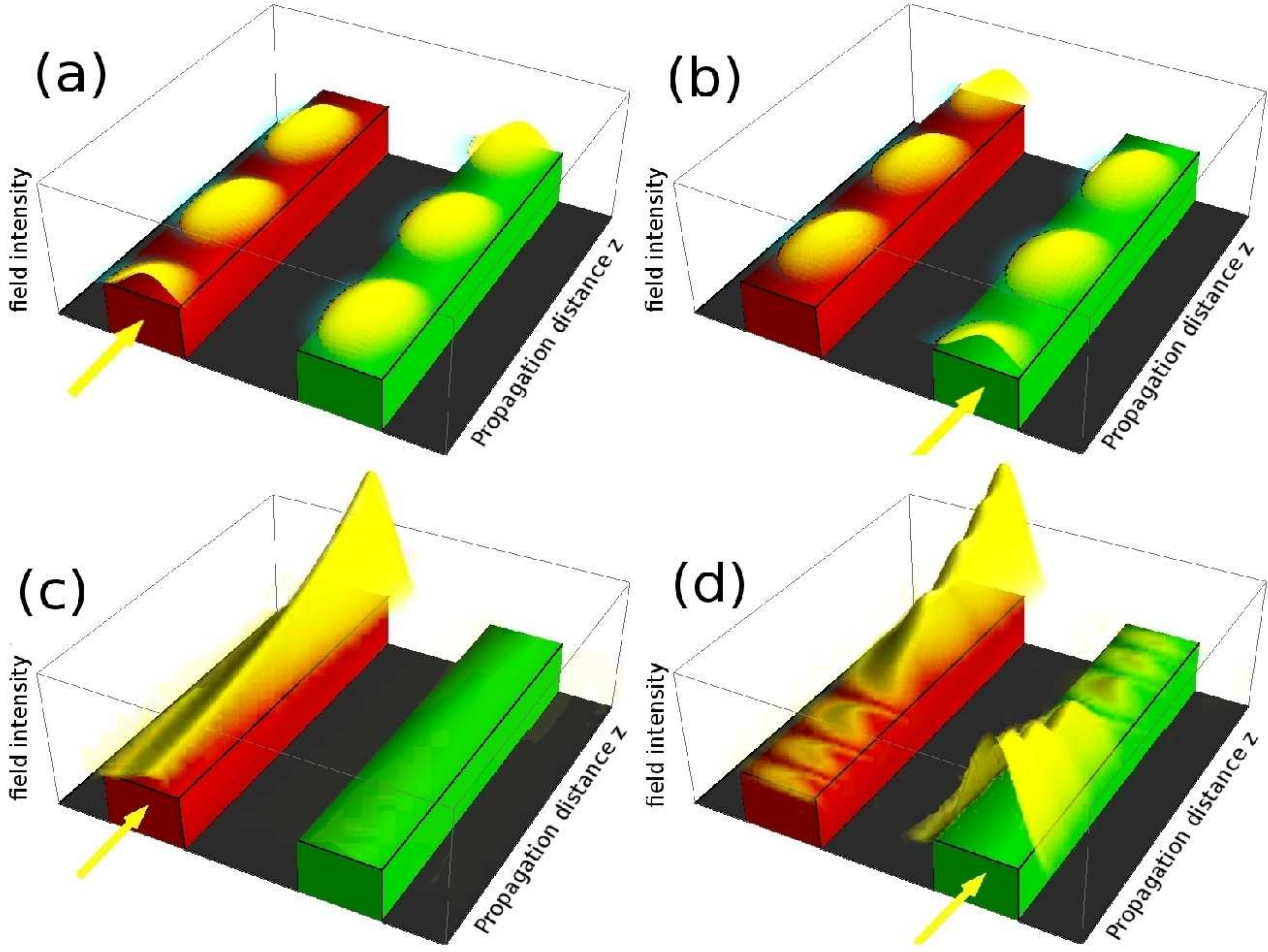}
 \caption{
   \label{fig1} 
   (color online) Beam propagation in two coupled nonlinear waveguides with non-linearity strength $\chi$ 
   and a complex ${\cal PT}$-symmetric refractive index profile. Waveguides are color-coded, indicating 
   balanced gain (red) and loss (green) regions ($\gamma=0.1$). Left columns correspond to an initial 
   excitation at the gain waveguide port, while right columns correspond to an initial excitation at the 
   lossy waveguide: (a,b) The non-linearity $\chi=1.9$ is below the critical value $\chi_d\approx 3.37$ 
   while for (c,d) the non-linearity strength $\chi=8$ is above.
   }
\end{figure}

For $\gamma=0$, Eqs. (\ref{eq5}) admits two constants of motion: the total input power $S_0$ and the 
total energy ${\cal H}=(\chi/2) S_3^2 + 2 S_1$. These two constants allow for an analytic solution
of the Stokes vector ${\overrightarrow S}$ in terms of elliptic functions \cite{J82}. Depending on 
the initial preparation and strength of nonlinearity $\chi$, we observe two distinct dynamical behaviors. 
For example, if the initial beam of total input power $S_0(0)=1$, is prepared in one of the two waveguides 
(i.e. $S_3(0)=\pm 1$), we observe either Rabi oscillations, or self-trapping dynamics \cite{J82}. 
The former case corresponds to $\chi<4$ and results in beam oscillations between the two waveguides, 
while the latter case, occurs for $\chi>4$ and leads to localization of the field (for all times) at 
the waveguide that was initially placed. In both cases, symmetric initial preparation will result 
to a dynamics which is reciprocal with respect to the axis of symmetry of the two coupled waveguides.

For $\gamma\neq 0$, neither the energy ${\cal H}$ nor the beam power $S_0$ are anymore conserved quantities.
Nevertheless, ${\cal PT}$-symmetry enforces two other constants of motion $C,J$: 
\begin{equation}
\label{eq6}
\begin{array}{cccc}
C^2&=&(\chi S_1-2)^2 + (\chi S_2)^2,&\quad \quad (a) \\ 
J&=&S_0 + {2\gamma \over \chi} \sin^{-1}\left({\chi S_1-2\over C}\right),&\quad\quad (b)
\end{array}
\end{equation}
thus indicating that the system of Eqs. (\ref{eq1}) is fully integrable. Below we will consider the 
case where initially $S_0(0)=1$, $S_3(0)=\pm 1$, while $S_1(0)=S_2(0)=0$. In this case, the constants 
of motion, as defined in Eqs. (\ref{eq6}), take the values $C_{\pm}=\pm 2$ and $J_{\pm}=1 \mp \gamma 
\pi/\chi$. 

Using $C$ and $J$, in this particular case we can express the components of the Stokes vector in terms of $S_0(z)$ in the 
following way
\begin{equation}
\label{eq7}
\begin{array}{cccc}
\chi S_1 & = & 2 \left( 1-\cos(\chi{1-S_0(z)\over 2\gamma}) \right),& (a) \\
\chi S_2 & = & 2 \sin(\chi{1-S_0(z)\over 2\gamma}), & (b)\\
\chi S_3 & = & \pm \sqrt{(\chi S_0(z))^2 - \left(4 \sin\left({\chi\over 4\gamma}(1-S_0(z)) \right)\right)^2}. & (c)\\
\end{array}
\end{equation}
Substituting the expression for $S_3$ from Eq. (\ref{eq7}c), to the first of the Eqs. (\ref{eq5}), we get 
that 
\begin{equation}
\label{eq7b}
\pm\int_{S_0=1}^{S_0(z)} {dS_0\over \sqrt{(\chi S_0)^2 - \left(4 \sin\left({\chi\over 4\gamma}(1-S_0)\right)\right)^2} } 
={2\gamma\over \chi} z. 
\end{equation}
Even though the problem is soluble by quadratures, the integral in Eq. (\ref{eq7b}) cannot be 
evaluated further and thus a closed expression for $S_0(z)$ is not possible. It is therefore 
instructive at this point to gain insight on the properties of the dynamics of this ${\cal PT}$ 
non-linear coupler by numerically solving Eqs. (\ref{eq1},\ref{eq5}). The 
accuracy of the numerical integration was checked via the conservation laws Eq. (\ref{eq6}), which 
were satisfied up to $10^{-10}$.

\begin{figure}[h]
\includegraphics[width=0.9\columnwidth,keepaspectratio,clip]{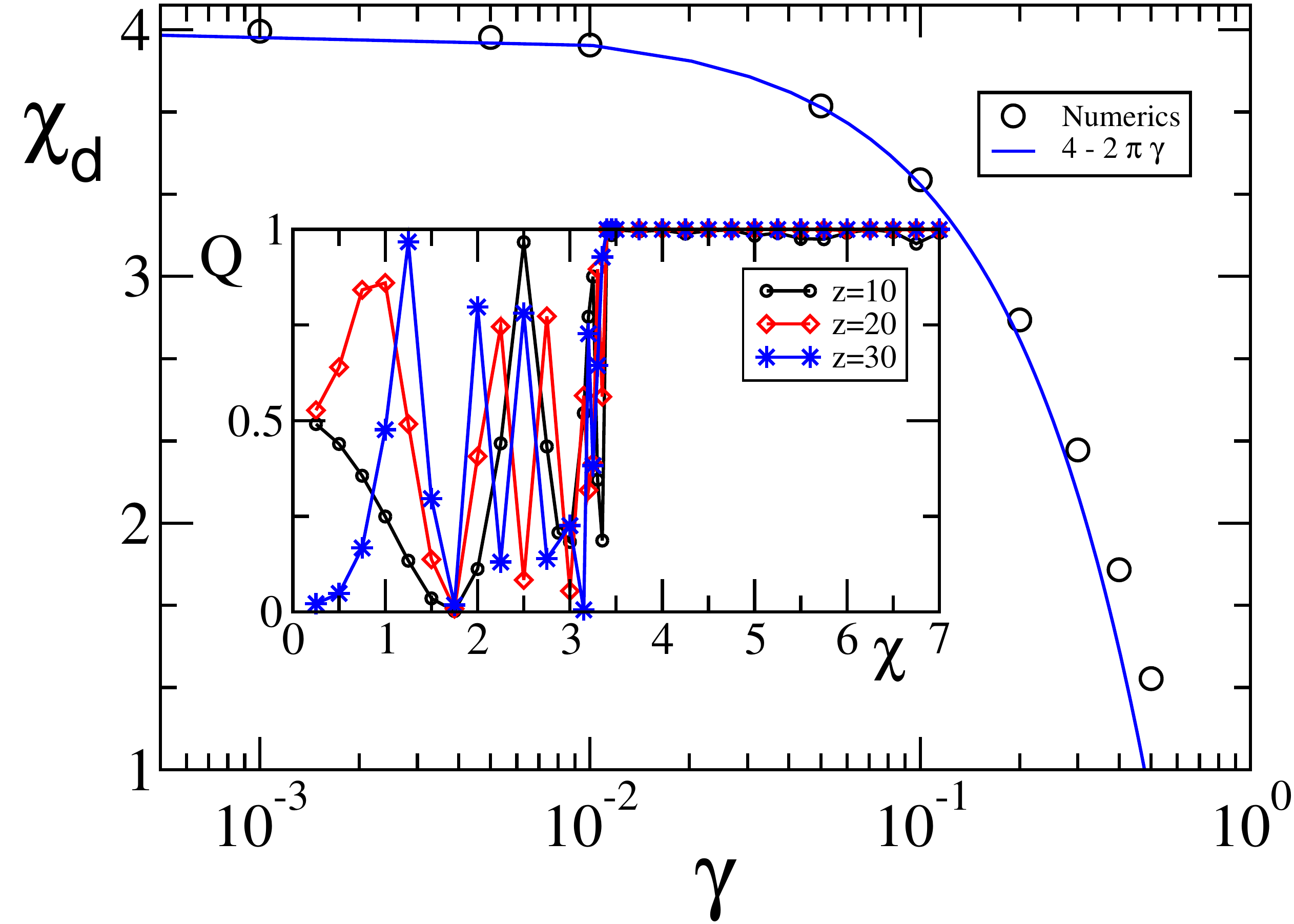}
 \caption{
   \label{fig2} 
(color online) Main figure: A semi-logarithmic plot of $\chi_d$ vs. $\gamma$. For the numerical evaluation
of $\chi_d$ we have integrated Eq. (\ref{eq1}) Inset: The efficiency factor $Q$ vs. nonlinearity strength $\chi$, for a 
fixed gain/loss parameter $\gamma=0.1$ and three different waveguide lengths $z=10,20$ and $z=30$. For
non-linearity strength $\chi=\chi_d\approx 3.4$ the isolator reaches its optimal efficiency.
   }
\end{figure}

Examples of the resulting beam dynamics for $\gamma =0.1$ and two representative non-linearity strengths 
$\chi=1.9$ and $\chi=8$ are reported in Fig. \ref{fig1}a,b and \ref{fig1}c,d 
respectively. In contrast to the $\gamma=0$ case \cite{J82}, now the dynamics is non-reciprocal with 
respect to the axis of symmetry of the system. While this is true for both values of non-linearity strength
$\chi$, it is much more pronounced for the case of Figs. \ref{fig1}c,d. In this latter case, the output 
field always leaves the sample from the waveguide with gain (red-colored) irrespective of the preparation
of the input beam. At the same time the output beam intensity at the lossy waveguide approaches zero for
longer waveguides. It is important to stress that in the case of the linear 
${\cal PT}$-dimer (see Figs 1e-f) the beam intensity at the lossy waveguide never goes to zero. Instead
it increases exponentially, albeit with a smaller prefactor with respect to the one of the gain waveguide.
This novel unidirectional propagation of the ${\cal PT}$-symmetric non-linear dimer is the key mechanism 
for establishing optical isolators (diodes). It has to be contrasted with the corresponding cases shown 
in Figs. \ref{fig1}a,b where the output beam depends on the input state, i.e. an initial excitation at 
the gain waveguide results in an output field at the lossy guide and vise versa. 

To quantify the ability of our set-up to act as an optical non-reciprocal device, we have defined the efficiency factor $Q$ 
of unidirectional propagation as
\begin{equation}
\label{effic}
Q(z)=1-\left| T_{+,+}(z) - T_{-,+}(z)\right|.
\end{equation}
where $T_{\pm,+}(z)\equiv |\psi_1(z)|^2/S_0(z)$ is the normalized transmission coefficient associated with 
the gain ($+$) waveguide of length $z$. In our definition we have always assumed that the initial input 
beam has total power $S_0(z=0)=1$, while the beam is launched either in the gain ($+$) or in the loss ($-$) 
waveguide. The efficiency factor takes values from $0\leq Q\leq 1$: a perfect diode corresponds to $Q=1$ (since 
the term inside the absolute value in Eq. (\ref{effic}) 
will be zero), while the opposite limit of $Q=0$ indicates total revival of the field. In the inset of Fig. 
\ref{fig2} we report our numerical findings for the efficiency factor $Q$ as a function of the non-linearity 
strength $\chi$ for three different waveguide lengths $z=10,20$, and $z=30$, and for a fixed value of the gain/
loss parameter $\gamma=0.1$. It is clear that an optimal diode is achieved once the non-linearity strength 
$\chi$ is larger than a critical value $\chi_d$.

\begin{figure}[h]
\includegraphics[width=0.9\columnwidth,keepaspectratio,clip]{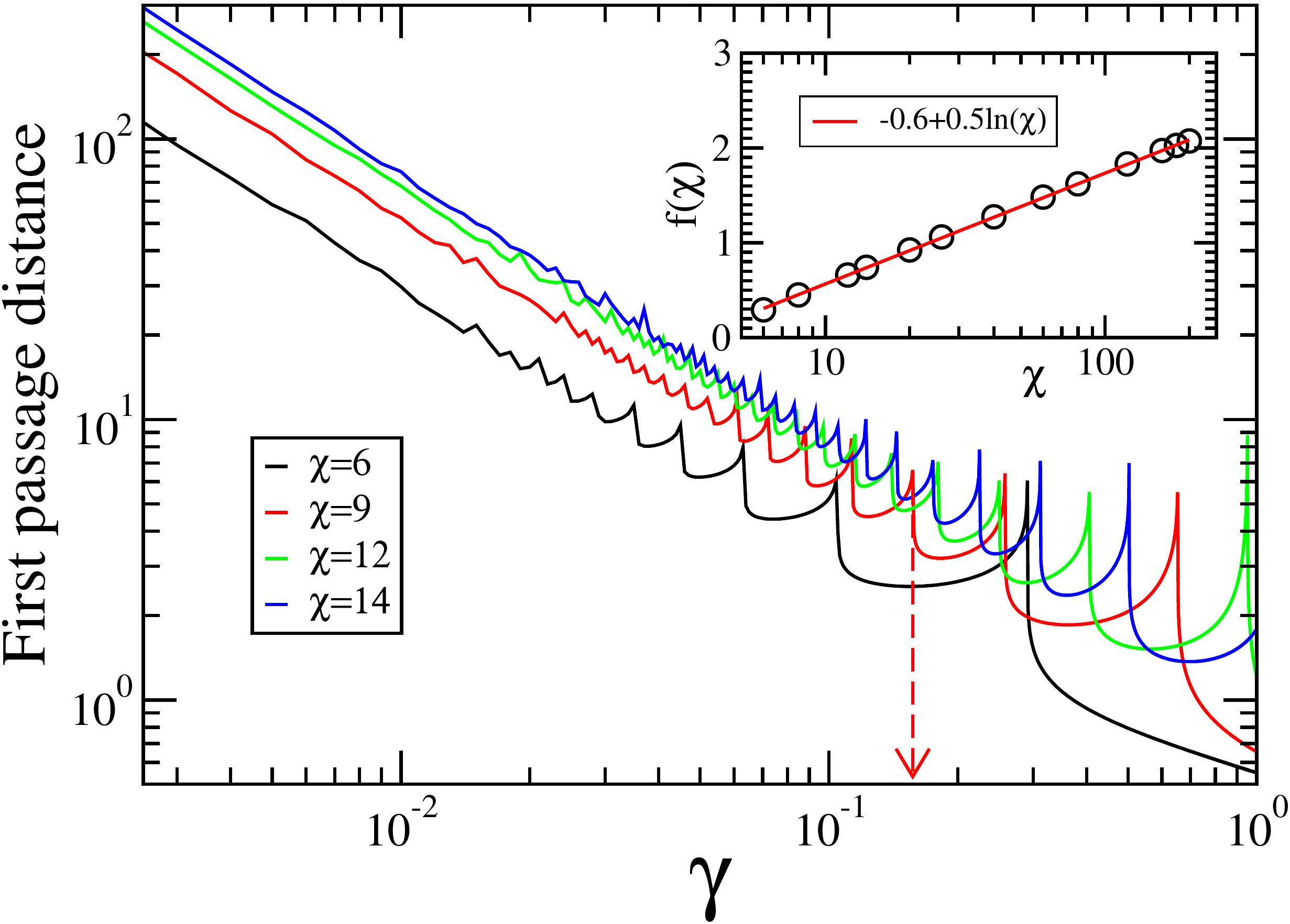}
 \caption{
   \label{fig3} 
(color online) The numerically extracted first passage distance $z_{\rm fpd}$ versus the gain/loss
parameter $\gamma$. The initial conditions are chosen to be $S_0(0)=1$ and $S_3(0)= -1$. An 
inverse power law is observed. Left inset: The proportionality coefficient $f(\chi)$ is plotted 
versus the nonlinearity strength $\chi$ for $\chi>\chi_d$. The red line correspond to the best 
linear fit. 
   }
\end{figure}

\subsection{Critical Non-linearity}

Next we present a heuristic argument that aims to estimate the critical non-linearity strength $\chi_d$ (as 
a function of $\gamma$), above which the ${\cal PT}$ symmetric non-linear dimer acts as an optical diode of
high efficiency factor $Q=1$. To this end we focus our analysis on the temporal behavior of the total power
$S_0(z)$. In the case of (Rabi-like) oscillations $S_0(z)$ is bounded between a minimum and a maximum value. 
Instead, in the regime where the coupled system acts as an optical diode, $S_0(z)$ is bounded only from below, 
while asymptotically it grows in an exponential fashion \cite{note1}. Using the first of Eq. (\ref{eq5}) together with Eq. 
(\ref{eq7}c), and requesting the extrema condition $dS_0(z)/dz=0$ (which is equivalent to $S_3(z)=0$) together 
with the condition $d^2S_0 /dz^2 <0$ for the existence of a global maxima, we find that $S_0(z)$ shows 
oscillatory behavior (i.e. Rabi-like oscillations) if the non-linearity $\chi$ is smaller than $\chi_d$, 
given by
\begin{equation}
\label{eq8}
\chi_{d}= 4 - 2 \pi \gamma
\end{equation} 
In the main panel of Fig.~\ref{fig2} we compare Eq. (\ref{eq8}) with the numerical values found for 
$\chi_d$. The latter has been evaluated via a direct integration of Eq. (\ref{eq1}) for systems
sizes up to $z=10^6$. The critical nonlinearity $\chi_d$ was evaluated up to a fourth digit accuracy
as the nonlinearity strength for which the total power $S_0(z)$, is bounded. In all cases the accuracy 
of the 
integration scheme has been guaranteed by requesting that the constants of motion Eq. (\ref{eq6}) are 
conserved with accuracy up to $10^{-5}$. A nice agreement between the theoretical and numerical value 
of $\chi_d$ is evident for small values of the gain/loss parameter $\gamma$, while 
deviations from the theoretical prediction start to be visible as $\gamma$ approaches the ${\cal PT}$ 
transition point (i.e. $\gamma=1$) of the linear system.

\begin{figure}[h]
\includegraphics[width=0.75\columnwidth,keepaspectratio,clip]{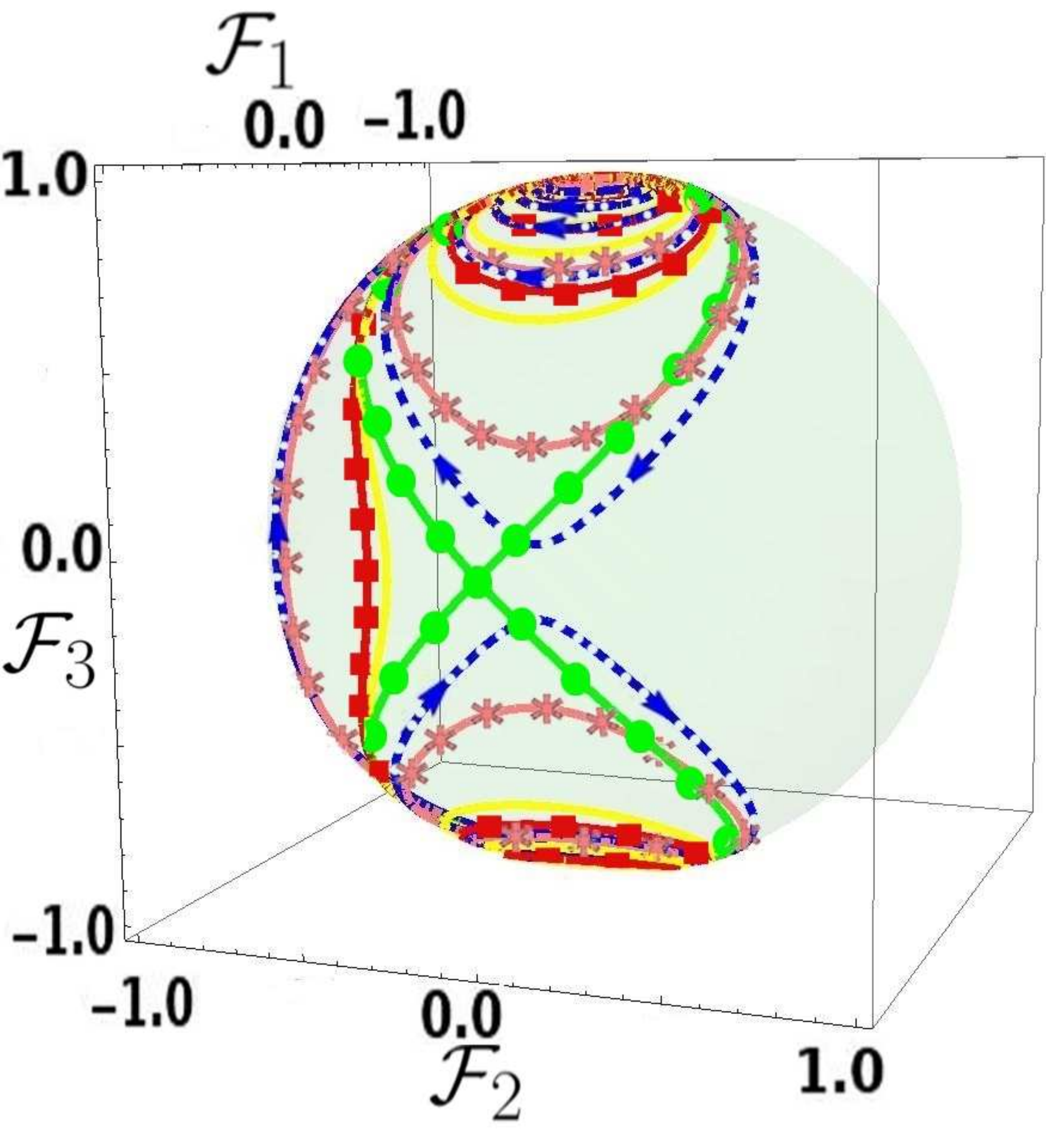}
 \caption{
   \label{fig4} 
(color online) Dynamics of the rescaled Stokes variables ${\cal F}$ for $\chi=9$ and various gain/loss parameters 
$\gamma$: dashed blue line correspond to $\gamma=0.157$; pink line ($\star$) correspond to $\gamma=0.15$; 
solid yellow line correspond to $\gamma =0.12$; and red line ($\blacksquare$) to $\gamma=0.174$. The green line 
($\bullet$) corresponds to the passive system $\gamma=0$ with critical non-linearity $\chi=4$, where the motion 
of the trajectory is on the separatrix. The trajectory associated with 
$\gamma=0.157$ (see red arrow in Fig. \ref{fig3}) is typical to the cases where $z_{\rm fpd}$ diverge and 
correspond to the closest one to the separatrix of the passive $\gamma=0$ system.
   }
\end{figure}

Finally, we investigate the minimal waveguide length $z_d$ which is required in order to have a high$-Q$ 
diode. From Figs. \ref{fig1}c,d we see that the beam evolution follow two distinct scenaria depending 
on the initial conditions: if the beam is launched initially at the gain waveguide, the propagation is 
mainly along this channel. If on the other hand the beam excites the lossy waveguide, there is a minimum 
propagation distance $z_d$ which is required before the light intensity 
is concentrated in the gain waveguide. We have found that $z_d$ is proportional to the "first passage 
distance" $z_{\rm fpd}$ associated with the point that $S_3$ becomes zero for the first time. In Fig. 
\ref{fig3} we report the results of our simulations for $z_d\sim z_{\rm fpd}$ for various $\chi (>
\chi_d)$ values or input power levels.

An intriguing feature of $z_{\rm fpd}$ is the existence of singularities (peaks in the $z_{\rm fpd}$) 
for some characteristic values of the gain/loss parameter $\gamma$. To understand the origin of these 
singularities, we have plotted the evolution of the Stokes vector ${\overrightarrow S}$, by making use 
of the rescaled variables $ {\overrightarrow {\cal F}}= {\overrightarrow S}/S_0$. In this representation, 
the magnitude $|{\overrightarrow {\cal F}}|$ remains constant, and thus we can visualize the evolution 
on the Bloch sphere (see Fig. \ref{fig4}). It should be emphasized that the Bloch trajectories can in 
general show self-intersections, as they are a projection from a higher dimensional phase space. One must 
also distinguish between closed orbits and those approaching an asymptotic state, as this is in general
connected to broken and unbroken ${\cal PT}$-symmetry \cite{BBM98}. We note that closely related Bloch 
dynamics appear in different physical model systems like the ones reported in \cite{other}. Our analysis,
indicated that the singularities in $z_{\rm fpd}$ are associated with trajectories that during their 
evolution, they stay close to the separatrix associated to the critical value $\chi=4$ (transition between 
Rabi-oscillations and self-trapping) of the passive system. 

Leaving aside the issue of the singularities, we have found that for all $\chi$-values larger 
than $\chi_d$, the first passage distance $z_{\rm fpd}$ follows an inverse power law i.e. 
\begin{equation}
\label{eq9}
z_{\rm fpd} = f(\chi)/\gamma
\end{equation}
where the proportionality factor $f(\chi)$ is $\chi$-dependent. A best least square fit allow us to 
extract the various $f(\chi)$ which is in this case $f(\chi)=-0.6+ 0.5 \ln(\chi)$
(see inset of Fig. \ref{fig3}).

\section{Conclusions}

In conclusion, we have proposed a new mechanism for directed transport in nonlinear optical coupled 
systems, that relies at the interplay between nonlinearity and ${\cal PT}$-reflection symmetries.
More specifically, we have observed that above a critical non-linearity strength, the beam evolution 
is unidirectional, i.e. the output beam remains in the gain channel, irrespective of initial conditions. 
Such behavior implies that these systems can be used to realize new 
classes of optical diodes and other unidirectional photonic elements. Of great interest will be to extend 
these notions to more involved arrangements like nonlinear ${\cal PT}$ lattices where nonlinear excitations 
are expected lead to even more intriguing phenomena.

We would like to acknowledge V. Kovanis for useful comments and suggestions. This work was supported by the DFG 
FOR760, and a grant from the US-Israel Binational Science Foundation (BSF), Jerusalem, Israel.



\begin{thebibliography}{99}
\bibitem{note1} In this case, the system of Eq. (\ref{eq1}) act as a set of two uncoupled waveguides i.e. in the long time
limit one can assume that the coupling constant is essentially zero. As a result the intensity at the gain waveguide (which is
approximately the same as the total intensity) increases exponentially, while the one at the lossy waveguide decays exponentially. 
We have checked the accuracy of this statement with direct numerical simulations.
\end{thebibliography}
\end{document}